\documentclass{article}

\usepackage[nonatbib,final]{neurips_2020}

\usepackage[utf8]{inputenc} % allow utf-8 input
\usepackage[T1]{fontenc}    % use 8-bit T1 fonts
\usepackage{hyperref}       % hyperlinks
\usepackage{url}            % simple URL typesetting
\usepackage{booktabs}       % professional-quality tables
\usepackage{amsmath}
\usepackage{amsfonts}       % blackboard math symbols
\usepackage{nicefrac}       % compact symbols for 1/2, etc.
\usepackage{microtype}      % microtypography
\usepackage{graphicx}
\usepackage{subcaption}

\title{DeepDFT: Neural Message Passing Network for Accurate Charge Density Prediction}
 
\author{%
  Peter Bjørn Jørgensen\\
  Energy Conversion and Storage\\
  Technical University of Denmark\\
  \And
  Arghya Bhowmik\\
  Energy Conversion and Storage\\
  Technical University of Denmark\\
  \texttt{arbh@dtu.dk} \\
}

\usepackage{bm}

\usepackage{xspace}		% As spaces after macros are ignored, we use xspace to
\usepackage{amssymb}	% Includes mathbb font among others
\usepackage{upgreek}
\usepackage{siunitx}

\renewcommand{\lim}[1]{\underset{#1}{\operatorname{lim}}\;}

\makeatletter
 \begingroup
 \catcode`\_=\active
 \protected\gdef_{\@ifnextchar|\subtextup\sb}
 \endgroup
 \def\subtextup|#1|{\sb{\textup{#1}}}
 \AtBeginDocument{\catcode`\_=12 \mathcode`\_=32768 }
\makeatother

\usepackage[draft]{fixme}
\newcommand{\figref}[1]{\figurename~\ref{#1}}

\begin{document}

\maketitle

\begin{abstract}
	We introduce DeepDFT, a deep learning model for predicting the electronic charge density around atoms $\rho(r)$, the fundamental variable in electronic structure simulations from which all ground state properties can be calculated.
	The model is formulated as neural message passing on a graph, consisting of interacting atom vertices and special query point vertices for which the charge density is predicted.
	The accuracy and scalability of the model are demonstrated for molecules, solids and liquids. The trained model achieves lower average prediction errors than the observed variations in charge density obtained from density functional theory simulations using different exchange correlation functionals.
  %The abstract paragraph should be indented \nicefrac{1}{2}~inch (3~picas) on
  %both the left- and right-hand margins. Use 10~point type, with a vertical
  %spacing (leading) of 11~points.  The word \textbf{Abstract} must be centered,
  %bold, and in point size 12. Two line spaces precede the abstract. The abstract
  %must be limited to one paragraph. $Arghya$
\end{abstract}

\section{Introduction}
Machine learning methods have been gaining popularity in materials research community recently \cite{butler2018machine}. Data driven models are being used to accelerate simulation based materials design through development of faster surrogate models that replace or work with physics based models.
In the realm of atomic scale modelling, the focus has been on the mapping of the molecular geometry to a target property, such as the total energy or the band gap \cite{noe2020machine}. Although more challenging to predict, the electronic charge density around atoms of a given molecule or material can describe the system completely, i.e. all ground state properties can be obtained from the charge density with reasonable accuracy \cite{cohen2012challenges}. It is the fundamental variable in Kohn Sham density functional theory (KS-DFT) \cite{payne1992iterative} based quantum simulations - the most popular electronic structure method used for materials research and design.
%In KS-DFT, an iterative self consistency process tries to converge to the actual charge density solution numerically.
DFT being an $O(n^3)$ method, computational cost inhibits simulating more than few hundred atoms. Computational design of materials from atomistic scale is severely limited by the system size as atomic scale representations of real materials and processes (e.g. battery electro-chemistry at cathode-electrolyte interface) needs $10^5$ atoms or more. In this work we aim at obtaining a linear scaling, general and flexible machine learning model that is able to utilise the vast amount of existing simulation results and those being produced daily in supercomputers around the world to predict the charge density for atomic structures of unprecedented size, enabling new avenues of atomic scale materials design. 

\section{State of the art and scope}

The first efforts towards predicting the electron density with machine learning \cite{brockherdeBypassingKohnShamEquations2017, bogojeski2018efficient} used a kernel ridge regression model, which takes, as input, an artificial Gaussian potential sampled on a 3D grid and outputs coefficients that represents the electron density in a Fourier basis set. The model can accurately predict the electron density in molecular dynamics simulations, however, by design, the model does not transfer across different molecular systems.
This limitation was later overcome with a similar kernel model \cite{grisafiTransferableMachineLearningModel2019, fabrizioElectronDensityLearning2019}, in which the total electron density is decomposed into additive atom-centered contributions and a symmetry-adapted Gaussian process regression model \cite{grisafi2018sagpr} is used to predict the expansion coefficients of each contribution based on the local environment around the atoms. The locality of the model allows it to transfer across different molecules, but the cubic computational cost (in number of training examples) of Gaussian process regression is prohibitive for applying the model to systems with a large number of atom types, where large number of data points are required to cover the chemical space of interest. Deep learning models generally scale well with the number of training examples and using these models might be the right direction for training with large data sets.
Among deep learning approaches - deep convolutional neural networks have been used to solve the charge density prediction problem by posing it as an image to image translation problem. For example, given a low accuracy DFT calculated density a neural network model predicts the output of high accuracy DFT \cite{sinitskiy2018deep}. Such an approach is restricted to a fixed voxel size and is not equivariant to rotations. A different approach is to manually construct fingerprints describing the local environment in 3D space and map the fingerprints to a density value using neural networks \cite{schmidtLearningModelsElectron2018a, chandrasekaranSolvingElectronicStructure2019, zepeda-nunezDeepDensityCircumventing2019}. Irrespective of machine learned or hand-engineered, features, when collected from longer distances (i.e. larger cutoff radius), may provide better accuracy. As the number of neighbours grow $O(r^3)$ with the cutoff radius $r$, a relatively small cutoff radius need to be used.
For neural message passing (or graph convolution) models, using a small graph connectivity cutoff in combination with multi-step message passing enables efficient propagation of long-range information in comparison to just increasing atomic environment cutoff distance - computational complexity grows only linearly with the number of message passing steps.

In the previous graph convolution based model \cite{gongPredictingChargeDensity2019} a new graph is created for each electron density query point. The atomic environment representation is thus dependent on the choice of query point and therefore the learned atomic environment representations do not directly transfer across different query points. In contrast, our model first obtain a representation of atoms and their local environment and in the next step map these representations to the electron density at the requested query points. This two-step approach allows the model to efficiently reuse the learned atomic environment representation.
Both steps are learned simultaneously in an end-to-end fashion.
In this work we demonstrate the excellent scalability and accuracy of our DeepDFT model. We train on very large scale datasets and achieve prediction errors lower than the variations in DFT charge density obtained with different exchange correlation (Xc) approximations.%better than DFT prediction accuracy.

%The novelty and high impact of the model lies in the usage of a smaller graph connectivity cutoff in combination with multi step message passing to achieve the similar effects of providing very long range structural information fingerprints directly. The representation learning provides exceptional scaling for large scale inference. In short we achieve both higher accuracy as well as lower computational complexity. %Furthermore, the mean average error of the ML-predicted charge densities is significantly higher 
%Why is this approach better
%using charge density makes property prediction less data intensive
%property prediction can be done through data/physics models once charge density is accessible 

\section{Neural Message Passing Network}
\begin{figure}[tp]
	\begin{minipage}{.45\linewidth}
		\centering\includegraphics[width=0.8\textwidth]{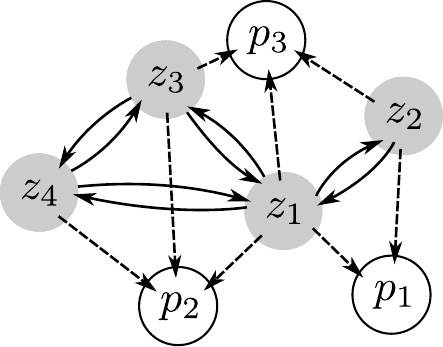}
		\caption{Messages are exchanged between atom vertices $z_i$ in several steps while the probe nodes $p_j$ only receive messages.}
		\label{fig:msgpassing_model}
	\end{minipage}\hfill%
	\begin{minipage}{.50\linewidth}
		\begin{picture}(180,120)
			%\put(0,0){\line(0,1){141}}
			%\put(0,0){\line(1,0){180}}
			\put(0,0){\includegraphics[width=180pt]{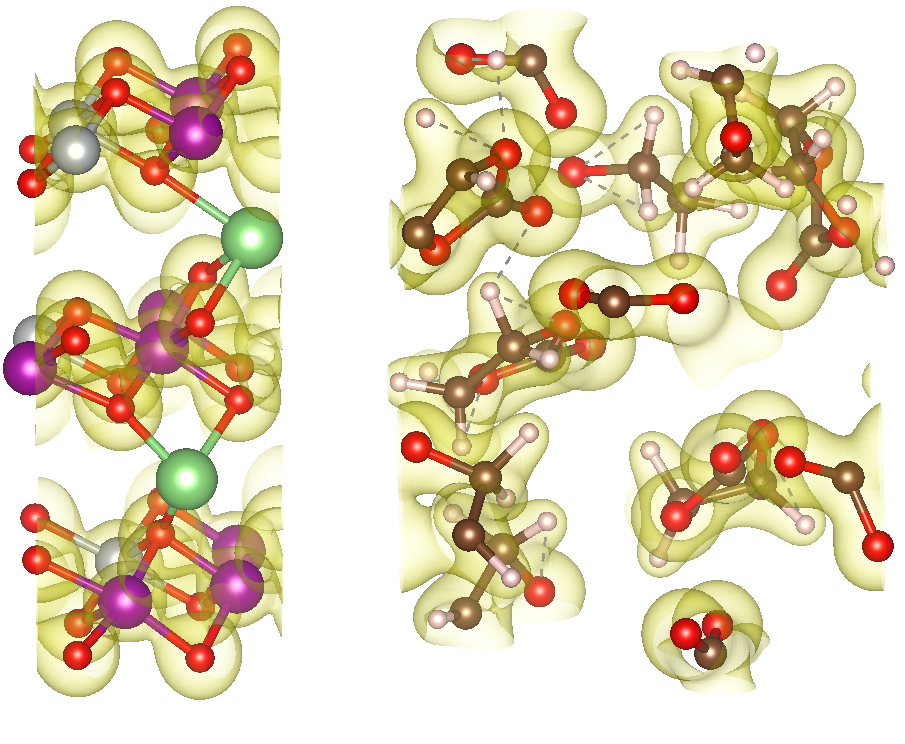}}
			\put(25,0){(a)}
			\put(125,0){(b)}
		\end{picture}
		%\centering\includegraphics[width=1.0\textwidth]{figs/data_examples_merged}
		\caption{Data set examples from (a) Lithium-Ion battery cathode - mixed transition metal (CO/Mn/Ni) layered oxide and (b) Ethylene-carbonate electrolyte in liquid state.}
	\end{minipage}%
\end{figure}

The DeepDFT density model is framed in the message passing framework devised by \cite{gilmerNeuralMessagePassing2017} and components of the model architecture is inspired by the SchNet model \cite{schuttSchNetDeepLearning2018}.
The input to the model is a graph representation of the molecule or crystal structure
and the graph has a vertex for each atom in the molecule or for each atom in the crystal structure unit cell. Edges are defined by a constant cutoff distance, i.e. we draw an edge between vertex $v$ and $w$ if the distance between $v$ and $w$ is less than a certain cutoff distance (\SI{4}{\angstrom}). The edges may cross periodic boundary conditions as in quotient graphs \cite{chungNomenclatureGenerationThreeperiodic1984, kleeCrystallographicNetsTheir2004}. Special probe vertices, that only accept incoming edges, are placed at each query point. A simple example graph with four atoms and three query points is illustrated in \figref{fig:msgpassing_model}.
%The SchNet model was not originally formulated in this framework, it was formulated as continuous convolution in 3D space with continuous radial filters.
%However as shown by \cite{jorgensen2018neural} the SchNet model also fits into this framework.

Each vertex $v$ has a hidden state $h_v^t$ at ``time'' step $t$. They are updated in a number of interaction ``time'' steps $T$.
We use $T=6$ interaction steps in all the experiments in this work.
%First we describe the atom-to-atom message passing model and then attend to the atom-to-query-point mapping.
The hidden states of the vertices are updated in two steps. First messages from neighbour vertices are computed using the message function $M_t(\cdot)$ and then the vertex state is updated by a state transition function $S_t(\cdot)$.
\begin{equation}
	m_{v}^{t+1} = \sum_{w \in N(v)} M_{t}(h_v^t, h_w^t, e_{vw}^{t+1}), \qquad \qquad
	h_v^{t+1} = S_t \left( h_v^t, m_{v}^{t+1} \right),
	\label{eq:updates}
\end{equation}
where $N(v)$ denotes the neighourhood of the vertex $v$.
For the initial atom vertex state we use the atomic number to look up an embedding vector for each vertex.
As edge feature we use the distance between the two corresponding atoms
expanded in a series of exponentiated quadratic functions:
\begin{align}
	(e_{vw}^0)_{k} = \exp \left(-\frac{(\varepsilon_{vw} - (-\mu_|min|+ k \Delta))^2}{2\Delta^2} \right),k=0\ldots k_|max|
	\label{eq:exponentiated}
\end{align}
where $\mu_|min|$, $\Delta$, and $k_|max|$ are chosen such that the centers of the functions covers the cutoff range.
This can be seen as a soft 1-hot-encoding of the distances, which makes it easier for a neural network to learn a function where the input distance is uncorrelated with the output of the network if that is necessary.
The atom-to-atom message function is a function of the sending vertex state $h_w$, the edge features $e_{vw}$, and can be written as
\begin{align}
	M_{t}(h_v^t, h_w^t, e_{vw}^t) &= M_{t}(h_w^t, e_{vw}^t) =  (W_{1}^{t} h_w^{t}) \odot c(d_{vw})g(W_{3}^{t}g(W_{2}^{t}e_{vw}^t)),
	\label{eq:message_function_impl}
\end{align}
where $\odot$ denotes element-wise multiplication, $g(\cdot)$ is the soft-plus activation function, and $c(\cdot)$ is a soft cutoff function $c(x)=1-\operatorname{sigmoid}(5*(x-(d_|cut|-1.5)))$ where $d_|cut|=\SI{4}{\angstrom}$ is the graph edge cutoff distance. %, i.e. when two atoms are within this distance they form a bond.
The expression on the right-hand side of the element-wise multiplication in \eqref{eq:message_function_impl} can be seen as a filter generating function \cite{schuttSchNetDeepLearning2018}.
%This is analogous to the discrete filters used in convolutional neural networks for image processing.
Because the filter depends only on the distance between atoms it becomes a radial filter rather than the directional filters normally used in image processing.

The state transition function is a two layer neural network on the sum of incoming messages and the result is added to the current hidden state as in Residual Networks \cite{he2015deep}:
\begin{equation}
	S_t \left( h_v^t, m_{v}^{t+1} \right) = h_v^t + W_5^t g( W_4^t m_{v}^{t+1}),
	\label{eq:state_transition}
\end{equation}

%To output a density we can probe any point in space. The probes are special vertices in the graph, they do not effect the state of the atom vertices hidden state or the state of any other probes, i.e. the probe vertices can only receive messages.
For the special probe vertices the hidden state is initialised as a vector of zeros.
We use the same form of message function as for the inter-atom messages \eqref{eq:message_function_impl}, but the weight matrices are not shared between the inter-atom message function and probe message function.
The state transition function for the probe vertices is also very similar to the atom hidden state transition function \eqref{eq:state_transition}. The only difference is that we add a ``forget'' gate term $F_t(\cdot)$, a two-layer neural network, that controls whether and which of the probe state entries that are updated or kept after each interaction step.
%The forget gate term is modeled as a two-layer neural network with a sigmoid activation function in the last layer.
The state transition functions for the probe vertices are thus given by:
\begin{align}
	\tilde{S}_t \left( h_v^t, m_{v}^{t+1} \right) &= F_t(h_v^t) \odot h_v^t + (1-F_t(h_v^t)) \odot W_5^t g( W_4^t m_{v}^{t+1}), \label{eq:probe_state_transition} \\
	\textrm{where } F_t \left( h_v^t  \right) &= \operatorname{sigmoid}(W_7^t g( W_6^t m_{v}^{t+1})),
\end{align}
The PyTorch implementation and pretrained model are available on Github.

\section{Dataset and Model Training}
To assess the model we use the QM9 dataset \cite{Ruddigkeit2012-dc, Ramakrishnan2014-ey} (134k small molecules with up to nine heavy atoms (CNOF)) that is widely used for benchmarking machine learning models for molecular property prediction. Additionally we also train and test with charge density data from crystalline and liquid state materials. 1:~A challenging to model but industrially important multi-transition  metal layered oxide lithium ion battery (LIB) cathode. 5000 configurations are generated through Monte Carlo style crystal site occupation for transition metal ions (Ni/Mn/Co) and lithium/vacancy to represent varieties of chemistry and lithiation state followed by structural optimization. 2:~Liquid ethylene carbonate - the most used LIB electrolyte. 12000 disordered configurations are generated through high temperature (3000K) accelerated identity preserving molecular dynamics. KS-DFT is used with PBE XC-functional and VASP code \cite{hafner2008ab} for obtaining valence charge density on a volumetric grid.
For each gradient step we sample 1000 query points in two training set molecules/crystal-structures and use the mean-squared-error at the query points as the cost function.
The model is trained for two weeks on a single Nvidia GPU and a small validation set is used for early stopping.

\section{Results}
\begin{figure}[tp]
	\centering\includegraphics[width=0.9\textwidth]{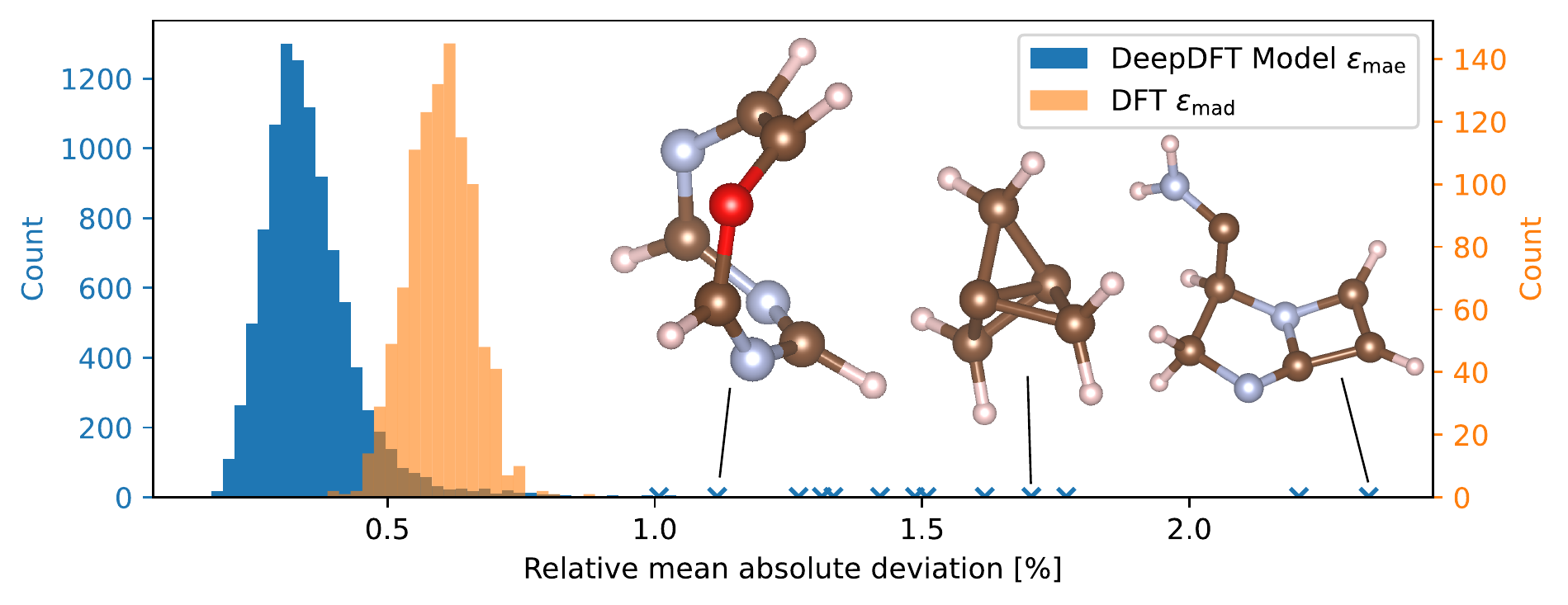}
	\caption{Histogram of relative MAE on QM9 test set. The markers show the outliers.}
	\label{fig:qm9_test_errors}
\end{figure}

To evaluate the model, we integrate the mean absolute error (MAE) over the whole simulation box normalized by total number of electrons (eq.~\eqref{eq:mae}) following \cite{grisafiTransferableMachineLearningModel2019, fabrizioElectronDensityLearning2019}. The DFT calculated density is used as ground truth $\rho(x)$.
The distribution of $\varepsilon_|mae|$ errors for the QM9 test set is shown in \figref{fig:qm9_test_errors}.
%Statistical uncertainty in DFT estimated electron density is larger than MAE of this model.
The DFT uncertainties, also included in the histogram, are estimated using eq.~\eqref{eq:mad} from an ensemble of densities $K=8$ calculated using VASP with eight different XC functionals (BF, PE, 91, CA, PZ, RP, RE, PS) \cite{wellendorff2012density, perdew1996generalized, perdew1992accurate, ceperley1980ground, perdew1981self, hammer1999improved, zhang1998comment, csonka2009assessing} for 1000 molecules. For each grid point we compute the mean absolute deviation (MAD) of the ensemble around the median density. The MAD is then integrated across all grid points and divided by the total number of electrons eq.~\eqref{eq:mad}.
%\vspace{-5mm}%
\noindent\begin{minipage}{.4\linewidth}
\begin{equation}
	\varepsilon_|mae| = \frac{\int_{x \in V} | \rho (x) - \hat{\rho} (x) |}{\int_{x \in V} |\rho(x) |}
	\label{eq:mae}
\end{equation}
\end{minipage}%
\begin{minipage}{.6\linewidth}
\begin{equation}
	\varepsilon_|mad| = \frac{\int_{x \in V} \frac{1}{K} \sum_{k=1}^K  | {\rho}_k (x) - \rho_|median| (x)  |}{\int_{x \in V} |\rho_1(x) |}
	\label{eq:mad}
\end{equation}
\end{minipage}

Predictions errors are mostly below \SI{0.5}{\percent} with an average of \SI{0.35}{\percent} and this is below the estimated DFT error which averages at \SI{0.6}{\percent}. This indicates our model is adept at learning from electronic structure simulations which are more accurate than KS-DFT. %, thus superseding KS-DFT at much lower computational cost.
For comparison, another baseline, which is the superposition of the single atom electron densities gives an average error of \SI{18}{\percent} for QM9 molecules (not shown). The DeepDFT model predictions include a number of outliers with very large errors compared to the average error, but they are still an order of magnitude more accurate than the superposition of single atom densities.
Manual inspection of the QM9 test set outliers shows that presence of exotic structural units (see \figref{fig:qm9_test_errors}), which are underrepresented or unrepresented in the training sets, leads to high error. For the other two datasets the average $\varepsilon_|mae|$ errors are also low (Table \ref{tab:datasets}) and no outliers are observed as chemical variations are well sampled. %This observation suggests that dataset creation based on molecular fingerprint similarity will enable higher accuracy with limited dataset. 

\begin{table}[tp]
  \caption{Datasets and prediction errors for DeepDFT model}
  \label{tab:datasets}
  \centering
  \begin{tabular}{lrrrSS}
    \toprule
    & \multicolumn{3}{c}{Dataset Splits} & \multicolumn{2}{c}{Test Set Error ($\varepsilon_|mae|$} \si{\percent})       \\
    \cmidrule(r){2-4}
    \cmidrule(r){5-6}
    Dataset  & {Train} & {Val.} & {Test} & {Average } & {Max.} \\
    \midrule
    QM9 & 123835   & 50 & 10000     &  0.36 & 2.35 \\
    LIB Cathode    & 3903 & 50 & 1000 &  0.10  & 0.29  \\
    Ethylene Carbonate &   7330    &  50 & 4000  & 0.53 & 0.69 \\
    \bottomrule
  \end{tabular}
\end{table}

\section{Conclusion}
%Achieving DFT level accuracy for systems like mixed transition metal oxides with defects, which are challenging for electronic structure methods due to their orbital complexity and disorder, opens up 
The proposed DeepDFT model achieves scalable and accurate prediction of electronic charge densities.
We envision that this type of model will allow design of new materials in a wide range of scientific/engineering problems where the size of the problem limits the applicability of conventional methods.
For example, in future work, we will investigate the capability of the model to capture charge transfers in reactive dynamical systems.
We will also investigate the data efficiency of the method, i.e. if learning the charge density first is more data efficient than directly predicting the properties of interest, such as total energy or band gaps.

%Achieving inexpensive machine learned density prediction, for large scale atomic structures with DFT level accuracy,  allows electronic structure based materials design in a wide variety of scientific/engineering problems where system size limits the applicability of conventional methods. We will also investigate the models' data efficiency, i.e. if learning the charge density first is more data efficient than directly predicting the properties of interest, such as total energy or band gaps. The effectiveness of the model in capturing charges transfer processes need to be tested for application in reactive dynamical systems. 

\section*{Broader Impact}

%Write something about releasing the model for the general public and enabling researchers around the world to come up with new breakthroughs.
The capacity to perform electronic structure simulations for thousands or even millions of atoms can potentially change how atomic scale simulation is used by scientists in the broad areas of biomedical sciences, engineering sciences and physical sciences. We want to provide democratic barrier free (in terms of the expertise, data and computational resources required) access to our trained models such that those can be deployed by non-experts without going though the data collection and training phase. Released software will be made user friendly such that experimental researchers without quantum physics background or even school students can perform electronic structure calculations without access to large computing resources.

Statistical models are trained on known phenomena and will make wrong predictions when used on systems for which the underlying mechanisms are very different and one might fail to discover these fundamentally new mechanisms if solely relying on statistical models. Thus, further research in machine learning uncertainty modeling is important in order to catch and avoid some of these pitfalls in the future. 

\begin{ack}
	Peter Bjørn Jørgensen and Arghya Bhowmik acknowledges financial support from VILLUM FONDEN by a research grant (00023105) for the DeepDFT project.
\end{ack}

\small

\bibliographystyle{unsrt}
\bibliography{neurips_2020.bib}

\begin{thebibliography}{10}

\bibitem{butler2018machine}
Keith~T Butler, Daniel~W Davies, Hugh Cartwright, Olexandr Isayev, and Aron
  Walsh.
\newblock Machine learning for molecular and materials science.
\newblock {\em Nature}, 559(7715):547--555, 2018.

\bibitem{noe2020machine}
Frank No{\'e}, Alexandre Tkatchenko, Klaus-Robert M{\"u}ller, and Cecilia
  Clementi.
\newblock Machine learning for molecular simulation.
\newblock {\em Annual review of physical chemistry}, 71:361--390, 2020.

\bibitem{cohen2012challenges}
Aron~J Cohen, Paula Mori-S{\'a}nchez, and Weitao Yang.
\newblock Challenges for density functional theory.
\newblock {\em Chemical reviews}, 112(1):289--320, 2012.

\bibitem{payne1992iterative}
Mike~C Payne, Michael~P Teter, Douglas~C Allan, TA~Arias, and ad~JD
  Joannopoulos.
\newblock Iterative minimization techniques for ab initio total-energy
  calculations: molecular dynamics and conjugate gradients.
\newblock {\em Reviews of modern physics}, 64(4):1045, 1992.

\bibitem{brockherdeBypassingKohnShamEquations2017}
Felix Brockherde, Leslie Vogt, Li~Li, Mark~E Tuckerman, Kieron Burke, and
  Klaus-Robert M{\"u}ller.
\newblock Bypassing the {{Kohn}}-{{Sham}} equations with machine learning.
\newblock {\em Nat. Commun.}, 8(1):872, October 2017.

\bibitem{bogojeski2018efficient}
Mihail Bogojeski, Felix Brockherde, Leslie Vogt-Maranto, Li~Li, Mark~E.
  Tuckerman, Kieron Burke, and Klaus-Robert Müller.
\newblock Efficient prediction of 3d electron densities using machine learning,
  2018.

\bibitem{grisafiTransferableMachineLearningModel2019}
Andrea Grisafi, Alberto Fabrizio, Benjamin Meyer, David~M Wilkins, Clemence
  Corminboeuf, and Michele Ceriotti.
\newblock Transferable {{Machine}}-{{Learning Model}} of the {{Electron
  Density}}.
\newblock {\em ACS Cent Sci}, 5(1):57--64, January 2019.

\bibitem{fabrizioElectronDensityLearning2019}
Alberto Fabrizio, Andrea Grisafi, Benjamin Meyer, Michele Ceriotti, and
  Clemence Corminboeuf.
\newblock Electron density learning of non-covalent systems.
\newblock {\em Chemical Science}, 10(41):9424--9432, 2019.

\bibitem{grisafi2018sagpr}
Andrea Grisafi, David~M. Wilkins, G\'abor Cs\'anyi, and Michele Ceriotti.
\newblock Symmetry-adapted machine learning for tensorial properties of
  atomistic systems.
\newblock {\em Phys. Rev. Lett.}, 120:036002, Jan 2018.

\bibitem{sinitskiy2018deep}
Anton~V. Sinitskiy and Vijay~S. Pande.
\newblock Deep neural network computes electron densities and energies of a
  large set of organic molecules faster than density functional theory (dft),
  2018.

\bibitem{schmidtLearningModelsElectron2018a}
Eric Schmidt, Andrew~T. Fowler, James~A. Elliott, and Paul~D. Bristowe.
\newblock Learning models for electron densities with {{Bayesian}} regression.
\newblock {\em Computational Materials Science}, 149:250--258, June 2018.

\bibitem{chandrasekaranSolvingElectronicStructure2019}
Anand Chandrasekaran, Deepak Kamal, Rohit Batra, Chiho Kim, Lihua Chen, and
  Rampi Ramprasad.
\newblock Solving the electronic structure problem with machine learning.
\newblock {\em npj Computational Materials}, 5(1):1--7, February 2019.

\bibitem{zepeda-nunezDeepDensityCircumventing2019}
Leonardo {Zepeda-N{\'u}{\~n}ez}, Yixiao Chen, Jiefu Zhang, Weile Jia, Linfeng
  Zhang, and Lin Lin.
\newblock Deep {{Density}}: Circumventing the {{Kohn}}-{{Sham}} equations via
  symmetry preserving neural networks.
\newblock {\em arXiv:1912.00775 [physics]}, November 2019.

\bibitem{gongPredictingChargeDensity2019}
Sheng Gong, Tian Xie, Taishan Zhu, Shuo Wang, Eric~R. Fadel, Yawei Li, and
  Jeffrey~C. Grossman.
\newblock Predicting charge density distribution of materials using a
  local-environment-based graph convolutional network.
\newblock {\em Physical Review B}, 100(18):184103, November 2019.

\bibitem{gilmerNeuralMessagePassing2017}
Justin Gilmer, Samuel~S Schoenholz, Patrick~F Riley, Oriol Vinyals, and
  George~E Dahl.
\newblock Neural {{Message Passing}} for {{Quantum Chemistry}}.
\newblock In {\em International {{Conference}} on {{Machine Learning}}}, pages
  1263--1272, July 2017.

\bibitem{schuttSchNetDeepLearning2018}
K~T Sch{\"u}tt, H~E Sauceda, P-J Kindermans, A~Tkatchenko, and K-R M{\"u}ller.
\newblock {{SchNet}} - {{A}} deep learning architecture for molecules and
  materials.
\newblock {\em J. Chem. Phys.}, 148(24):241722, June 2018.

\bibitem{chungNomenclatureGenerationThreeperiodic1984}
S~J Chung, Th~Hahn, and W~E Klee.
\newblock Nomenclature and generation of three-periodic nets: The vector
  method.
\newblock {\em Acta Crystallogr. A}, 40(1):42--50, January 1984.

\bibitem{kleeCrystallographicNetsTheir2004}
W~E Klee.
\newblock Crystallographic nets and their quotient graphs.
\newblock {\em Cryst. Res. Technol.}, 39(11):959--968, November 2004.

\bibitem{he2015deep}
Kaiming He, Xiangyu Zhang, Shaoqing Ren, and Jian Sun.
\newblock Deep residual learning for image recognition, 2015.

\bibitem{Ruddigkeit2012-dc}
Lars Ruddigkeit, Ruud van Deursen, Lorenz~C Blum, and Jean-Louis Reymond.
\newblock Enumeration of 166 billion organic small molecules in the chemical
  universe database {GDB-17}.
\newblock {\em J. Chem. Inf. Model.}, 52(11):2864--2875, November 2012.

\bibitem{Ramakrishnan2014-ey}
Raghunathan Ramakrishnan, Pavlo~O Dral, Matthias Rupp, and O~Anatole von
  Lilienfeld.
\newblock Quantum chemistry structures and properties of 134 kilo molecules.
\newblock {\em Sci Data}, 1:140022, August 2014.

\bibitem{hafner2008ab}
J{\"u}rgen Hafner.
\newblock Ab-initio simulations of materials using vasp: Density-functional
  theory and beyond.
\newblock {\em Journal of computational chemistry}, 29(13):2044--2078, 2008.

\bibitem{wellendorff2012density}
Jess Wellendorff, Keld~T Lundgaard, Andreas M{\o}gelh{\o}j, Vivien Petzold,
  David~D Landis, Jens~K N{\o}rskov, Thomas Bligaard, and Karsten~W Jacobsen.
\newblock Density functionals for surface science: Exchange-correlation model
  development with bayesian error estimation.
\newblock {\em Physical Review B}, 85(23):235149, 2012.

\bibitem{perdew1996generalized}
John~P Perdew, Kieron Burke, and Matthias Ernzerhof.
\newblock Generalized gradient approximation made simple.
\newblock {\em Physical review letters}, 77(18):3865, 1996.

\bibitem{perdew1992accurate}
John~P Perdew and Yue Wang.
\newblock Accurate and simple analytic representation of the electron-gas
  correlation energy.
\newblock {\em Physical review B}, 45(23):13244, 1992.

\bibitem{ceperley1980ground}
David~M Ceperley and Berni~J Alder.
\newblock Ground state of the electron gas by a stochastic method.
\newblock {\em Physical Review Letters}, 45(7):566, 1980.

\bibitem{perdew1981self}
John~P Perdew and Alex Zunger.
\newblock Self-interaction correction to density-functional approximations for
  many-electron systems.
\newblock {\em Physical Review B}, 23(10):5048, 1981.

\bibitem{hammer1999improved}
Bj{\o}rk Hammer, Lars~Bruno Hansen, and Jens~Kehlet N{\o}rskov.
\newblock Improved adsorption energetics within density-functional theory using
  revised perdew-burke-ernzerhof functionals.
\newblock {\em Physical review B}, 59(11):7413, 1999.

\bibitem{zhang1998comment}
Yingkai Zhang and Weitao Yang.
\newblock Comment on “generalized gradient approximation made simple”.
\newblock {\em Physical Review Letters}, 80(4):890, 1998.

\bibitem{csonka2009assessing}
G{\'a}bor~I Csonka, John~P Perdew, Adrienn Ruzsinszky, Pier~HT Philipsen,
  S{\'e}bastien Leb{\`e}gue, Joachim Paier, Oleg~A Vydrov, and J{\'a}nos~G
  {\'A}ngy{\'a}n.
\newblock Assessing the performance of recent density functionals for bulk
  solids.
\newblock {\em Physical Review B}, 79(15):155107, 2009.

\end{thebibliography}

\end{document}